\documentclass[aps,prb,twocolumn,groupedaddress,showpacs,floatfix,altaffilletter,longbibliography,includegraphics]{revtex4-2}
\usepackage{graphicx}
\usepackage{grffile}
\usepackage{amsmath}
\usepackage{amssymb}
\usepackage{bm}
\usepackage{color}
\usepackage[dvipsnames]{xcolor}
\usepackage{amsmath,amssymb,amsfonts}
\usepackage{epsfig}
\usepackage{times}
\usepackage[colorlinks,bookmarks=false,citecolor=blue,linkcolor=red,urlcolor=blue]{hyperref}
\usepackage{gensymb}
\usepackage[toc,page]{appendix}
\usepackage{float}
\usepackage{comment}

\newcommand{\secref}[1]{Sec.~(\ref{#1})}

\newcommand{\fref}[1]{Fig.~\ref{#1}}
\newcommand{\eref}[1]{Eq.~(\ref{#1})}

\begin{document}

\title{Antiferromagnetic pseudogap in the two-dimensional Hubbard model deep in the renormalized classical regime}

\author{ Y.M. Vilk}
\affiliation{ 33 Weatherly Dr, Salem, MA 01970}
\author{Camille Lahaie$^{1}$, and A.-M.S.~Tremblay$^{1}$}
\affiliation{$^1$D\'epartement de physique \& Institut quantique \& RQMP\\
Universit\'e de Sherbrooke, Sherbrooke, Qu\'ebec, J1K 2R1 Canada}
\date{\today}
\begin{abstract}
Long-wavelength spin fluctuations prohibit antiferromagnetic long-range order at finite temperature in two dimensions. Nevertheless, the correlation length starts to grow rapidly at a crossover temperature, leading to critical slowing down and to a renormalized-classical regime over
 a wide range of temperature, between a fraction of the mean-field transition temperature and the zero-temperature ordered state. 
This leads to a single-particle pseudogap of the kind observed in electron-doped cuprates. Very few theoretical methods can claim an accurate description of this regime. The challenge is that in this regime Fermi-liquid quasiparticles are already destroyed and new quasiparticles of the ordered state are not fully formed yet.
Here, we study this problem for the two-dimensional Hubbard model by first generalizing the two-particle self-consistent approach. 
Using a special algorithm, spin fluctuations are treated in the thermodynamic limit even for large correlation lengths. The effects of Kanamori-Br\"uckner screening, of classical and of quantum fluctuations are taken into account. 
Results are presented at half-filling for the one-band Hubbard model with nearest-neighbor hopping.  They agree well with available benchmark diagrammatic quantum Monte Carlo at high temperature where the pseudogap opens up. In addition to temperature-dependent spectral properties, we find quantum corrections to the zero-temperature renormalized mean-field antiferromagnetic gap. Finally, analytic continuation of the Matsubara results for spectral functions show that the pseudogap opens up at significantly higher temperature than was previously identified based on the Matsubara data only.  
\end{abstract}

\maketitle

% Introduction
\section{Introduction}
\label{sec:intro}

The pseudogap in the high temperature superconductors was observed in both hole-doped \cite{ding_spectroscopic_1996,loeser_excitation_1996,renner_pseudogap_1998,timusk1999pseudogap,vishik2018photoemission} and electron-doped materials \cite{armitage2001anomalous,motoyama_spin_2007}. Many theories were proposed for explanation of the pseudogap \cite{kampf1990pseudogaps,Vilk1996, Vilk1997,varma1997non,norman1998phenomenology,franz1998phase,chen1998pairing,schmalian_microscopic_1999,senechal2004hot,kyung2006pseudogap,  Ferrero_Cornaglia_De_Leo_Parcollet_Kotliar_Georges_2009, Sordi_Haule_Tremblay_2010,  Gull_Ferrero_Parcollet_Georges_Millis_2010, sordi2012pseudogap, Gull_Millis_2013,  
Sordi_Semon_Haule_tremblay_2013, atkinson2015charge, 
 Fratino_Sémon_Sordi_Tremblay_2016, 
 Wu_Scheurer_Chatterjee_Sachdev_Georges_Ferrero_2018, 
 wu2021interplay, Fratino_Bag_Camjayi_Civelli_Rozenberg_2021, gauvin2022disorder, Walsh_Charlebois_Sémon_Sordi_Tremblay_2022, wang2023phase, ye_location_2023, dai2020modeling,Ye_Chubukov_2023,sakai2023nonperturbative}. There is no current consensus on the nature of the pseudogap. One of the discussed mechanisms for the pseudogap is critical antiferromagnetic fluctuations. In the hole-doped materials the antiferromagnetic correlation length is short, it does not grow fast with decreasing temperature and the pseudogap occurs above the superconducting state. It thus seems {\it unlikely} that the origin of the pseudogap in the hole-doped high temperature superconductors is due to  long wavelength antiferromagnetic fluctuations. 
 
By contrast, in the the electron-doped high temperature superconductors, the pseudogap was observed in the regime where the antiferromagnetic correlation length grows exponentially \cite{armitage2001anomalous,motoyama_spin_2007}. It was shown~\cite{Kyung2004,motoyama_spin_2007} that the pseudogap in these quasi-two-dimensional materials can be explained quantitatively by  critical antiferromagnetic fluctuations in the renormalized classical regime.

 The qualitative explanation is as follows. In two dimensions(2D), due to the Mermin-Wagner theorem the mean-field phase transition to the antiferromagnetic state is suppressed by critical spin fluctuations. 
 This results in a wide temperature range where the antiferromagnetic correlation length grows exponentially. In this regime, Fermi-liquid quasiparticles are already destroyed and new quasiparticles of the ordered state are not fully formed yet, leading to a single-particle pseudogap that is a precursor of the zero-temperature antiferromagnetic gap.  
 
 Theoretically, this pseudogap in  weak to intermediate coupling has been predicted in the context of the repulsive 2D Hubbard model at half-filling using the two-particle self-consistent approximation (TPSC) \cite{Vilk1996, Vilk1997}. 
 The TPSC approach satisfies the Pauli principle, the local spin and charge sum rules and the Mermin-Wagner theorem and predicts a wide renormalized classical (RC) regime in which the pseudogap exists. 
 That work also established the criterion for the pseudogap: the pseudogap appears when the correlation length $\xi$ for spin fluctuations exceeds the thermal de Broglie wavelength of  electrons $\xi_{th}=v_F/\pi T$ (in units where $\hbar=1$ and $k_B=1$). 
 Here $v_F$ is an electron velocity at the Fermi surface and $T$ is the temperature
 \footnote{This criterion can be modified in the presence of disorder~\cite{gauvin2022disorder}}.
 The TPSC approach has been extended to  more realistic multi-orbital models \cite{Miyahara_2013, Zantout_2021, Gauvin-Ndiaye_Leblanc_Marin_Martin_Lessnich_Tremblay_2024}, to spin-orbit coupling~\cite{lessnich2024spin}, to out-of-equilibrium calculations \cite{Simard_2022}, to the attractive Hubbard model \cite{vilk1998attractive} and to disorder effects \cite{gauvin2022disorder}. The flexibility of this method and its low computational cost makes it attractive for applications to real materials.

 Theoretical studies of the antiferromagnetic pseudogap were extended in a number of directions:  $t-t'$ model in the presence of doping\cite{vilk1997shadow,schmalian_weak_1998,schmalian_microscopic_1999,kuchinskii_models_1999}, quasi-two-dimensional system \cite{dare1996crossover, moca_analytical_2000} and approximate methods that go beyond one-loop approximation for the self-energy \cite{sedrakyan_pseudogap_2010,Ye_Chubukov_2023,nikolaenko_spin_2023}. 

 The existence of the pseudogap in the 2D Hubbard model at half-filling has been confirmed by statistically exact numerical calculations with both diagrammatic (DiagMC) \cite{vsimkovic2020extended,Schaefer2021} and determinantal (DQMC) quantum Monte Carlo \cite{Moukouri2000,schafer2015fate,Schaefer2021}. 
 These methods convincingly show that, in this model, the pseudogap develops in the renormalized classical regime. 
 While these numerical methods work well in the metallic state and when the pseudogap begins to open-up, they run into convergence problems at low temperatures when the spin correlation length becomes large. 
 They are also computationally expensive and difficult to generalize to the more realistic three-band models. 
 Hence we go back to the TPSC approach to apply it to regimes that are presently inaccessible to exact methods.

However, while the TPSC approach agrees quantitatively with benchmarks Monte Carlo results at high temperatures~\cite{Vilk1997,TremblayMancini:2011,Moukouri2000} and gives a  description of the crossover to the renormalized classical regime, it fails deep inside the pseudogap regime where it leads to some nonphysical results. 
For example, at half-filling in the nearest-neighbor hopping Hubbard model, it predicts that double occupancy decreases to zero at zero temperature. 
The modified TPSC+ approach proposed in Ref.~\cite{gauvin2023improved} remedies this by taking into account the feedback of the single-particle properties on the spin and charge susceptibilities. 
TPSC+ provides a framework for studying strongly correlated electron systems deep inside the pseudogap regime where no well-defined quasiparticles exist. 
The approach also improved agreement with benchmarks in the metallic regime. However, due to  technical convergence issues, the study in  Ref.~\cite{gauvin2023improved} was limited to  temperatures above the pseudogap crossover defined by DiagMC results. 
   
Here, we overcome this limitation and  obtain results in the thermodynamic limit (TDL)  for practically any temperature and spin correlation lengths up to $\xi \sim 10^6 $ (we take lattice spacing as unity). 
We also refine this approach further and obtain good agreement with benchmark DiagMC results at the beginning of the pseudogap regime {\it at half-filling} for the one-band Hubbard model with nearest-neighbor hopping.  
We also estimate the zero-temperature gap by reaching very low temperatures and showing that it  significantly deviates from the mean-field results (Stoner criteria) due to quantum fluctuations. 
Finally, we perform analytical continuation of the single-particle properties using Padé approximation: 
We find that the pseudogap in the spectral function opens up at significantly higher temperature than was determined earlier using Matsubara data only.

	The paper is organized as follows. 
 In \secref{sec:model}, we discuss the model and review the TPSC and TPSC+ equations. 
 In \secref{sec:results} we compare our results with DiagMC in the pseudogap regime and estimate the zero-temperature gap.  
 We also show results for the analytical continuation of the spectral function and the density of states. 
 In Appendix~\ref{app:Num_Impl} we explain the numerical implementation, including the thermodynamic limit and how convergence is achieved. 
	
\section{Model and theoretical approach}
\label{sec:model}

After introducing the model, we recall the TPSC approach, what it teaches us about the pseudogap and why TPSC fails in the renormalized classical regime. We then move to the TPSC+ approach and introduce a slight variant, TPSC+Classical. 

\subsection{The Hubbard Model}
We study the one-band Hubbard model on a square $2D$ lattice at half-filling, 
\begin{multline}
H = - t\sum_{\langle ij\rangle,\sigma}(c^{\dagger}_{i,\sigma} c_{j,\sigma} + c^{\dagger}_{j,\sigma} c_{i,\sigma})\\
 + U\sum_{i} n_{i,\uparrow} n_{i, \downarrow} - \mu\sum_{i, \sigma} n_{i,\sigma}.
\end{multline}
The nearest-neighbor hopping $t$ is our unit of energy, $U$ is the Hubbard repulsion on a single site, $\mu$ the chemical potential, $n_{i\sigma}$ counts the number of electrons of spin $\sigma$ at site $i$ with the corresponding annihilation (creation) operators  $c^{(\dagger)}_{i,\sigma}$. The units are $\hbar=1$, $k_B=1$, lattice spacing $a=1$. In these units the dispersion is $\epsilon_\mathbf{k}=-2(\cos(k_x)+\cos(k_y))$ where $\mathbf{k}$ is wave vector. 

\subsection{The two-particle self-consistent (TPSC) approach}

In this subsection, we describe some of the main properties of the TPSC approach. This nonperturbative method satisfies the Pauli principle, the Mermin-Wagner theorem and conservation laws for the spin and charge susceptibilities. The spin and charge susceptibilities have RPA-like forms but with bare interaction $U$ replaced by renormalized spin $U_{sp} $ and charge $U_{ch} $ interactions 

%%%%%%%%%%%%%%%%
\begin{equation}
    \chi_{sp}(\mathbf{q},iq_n) = \frac{\chi^{(1)}(\mathbf{q},iq_n)}{1-\frac{U_{sp}}{2}\chi^{(1)}(\mathbf{q},iq_n)},
    \label{eq:chispqiqn}
\end{equation}
%%%%%%%%%%%%%%%%

%%%%%%%%%%%%%%%%
\begin{equation}
    \chi_{ch}(\mathbf{q},iq_n) = \frac{\chi^{(1)}(\mathbf{q},iq_n)}{1+\frac{U_{ch}}{2}\chi^{(1)}(\mathbf{q},iq_n)}.
    \label{eq:chichqiqn}
\end{equation}
%%%%%%%%%%%%%%%%
 Here $q_n = 2n\pi T$  are bosonic Matsubara frequencies while $\mathbf{q}$ are wave vectors in the Brillouin zone. The bubble $\chi^{(1)}$, evaluated at the first level of approximation, is 
%%%%%%%%%%%%%%%%
\begin{equation}
    \chi^{(1)}(\mathbf{q},iq_n) = -2\frac{T}{N}\sum_{\mathbf{k},ik_n}\mathcal{G}^{(1)}_\sigma(\mathbf{k},ik_n)\mathcal{G}^{(1)}_\sigma(\mathbf{k}+\mathbf{q},ik_n+iq_n),
    \label{eq:chi1qiqn}
\end{equation}
%%%%%%%%%%%%%%%%
where $N$ is the total number of sites in the system, $T$ is the temperature, $k_n=(2n+1)\pi T$ are fermionic Matsubara frequencies and $\mathbf{k}$ are wave vectors in the first Brillouin zone.

 The susceptibility $\chi^{(1)}(q)$ is identical to the noninteracting case but it is important to keep in mind that the self-energy entering the Green's function $\mathcal{G}^{(1)}_{\sigma}(k)$ at this level is a constant which can be absorbed into the chemical potential for a given filling $n$ \cite{TremblayMancini:2011}. 

The renormalized vertices $U_{sp} $ and $U_{ch} $ are found self-consistently using the local spin and charge sum rules~\cite{vilk1994theory}:
%%%%%%%%%%%%%%%
\begin{align}
\frac{T}{N}\sum_{\mathbf{q},q_n} \chi_{sp}(\mathbf{q},iq_n) &= n-2 \langle n_{\uparrow}n_{\downarrow} \rangle, \label{eq:sumrule_sp}\\
\frac{T}{N}\sum_{\mathbf{q},q_n} \chi_{ch}(\mathbf{q},iq_n) &= n+2 \langle n_{\uparrow}n_{\downarrow} \rangle - n^2.  \label{eq:sumrule_ch}
\end{align}
%%%%%%%%%
 Note that to arrive at the expressions on the right one needs to use the Pauli principle $n_{\sigma}^2=n_{\sigma}$. 

To compute self-consistently the vertices $U_{sp}$ and $U_{ch}$ from the sum rules, a third equation is needed. In  TPSC, the following \emph{ansatz}~\cite{vilk1994theory,hedayati1989ground} for  $U_{sp}$ is used:
%%%%%%%%%%%
\begin{equation}
U_{sp} = U \frac{\langle n_{\uparrow}n_{\downarrow} \rangle}{\langle n_{\uparrow} \rangle \langle n_{\downarrow} \rangle}  \label{eq:ansatz_tpsc}
\end{equation}
%%%%%%%%%%%%%%
The renormalized vertex $U_{sp}$ is reduced relative to the bare interaction by the pair correlation function $g_{\uparrow \downarrow} (0) = \langle n_{\uparrow} n_{\downarrow} \rangle/ \langle n_{\uparrow} \rangle  \langle n_{\downarrow} \rangle$. The Hartree-Fock approximation corresponds to the noninteracting value  $g_{\uparrow \downarrow} (0)=1$ and the reduction of $g_{\uparrow \downarrow} (0) $ from unity describes Kanamori-Brückner screening. The latter is responsible for the absence of the ferromagnetic order in the Hubbard model \cite{vilk1994theory}.  

In analogy with the procedure in the electron gas, the self-energy is computed in a second step, using the spin and charge susceptibilities from the first step that contain collective modes \cite{Vilk1996,Vilk1997,Moukouri2000}:

\begin{align}
    \Sigma_\sigma^{(2)}(\mathbf{k},&ik_n) = Un_{-\sigma}\nonumber+\frac{T}{N} U\sum_{\mathbf{q},iq_n}\left [\tilde{g} U_{sp}\chi_{sp}(\mathbf{q},iq_n)\right.\nonumber \\
     &+\left.(1/2 - \tilde{g}) U_{ch}\chi_{ch}(\mathbf{q},iq_n)\right ] \mathcal{G}_\sigma^{(1)}(\mathbf{k}+\mathbf{q},ik_n+iq_n).		
    \label{eq:selfEnergy2}
\end{align}   
Contrary to the electron-phonon case, there is no Migdal theorem. Spin and charge vertices are renormalized.

 In the original TPSC paper \cite{Vilk1996}, \cite{Vilk1997} the value $\tilde{g} =1/4$ was used and in a later work \cite{Moukouri2000} it was argued that  $\tilde{g} =3/8$ was a better choice. The ambiguity is due to the fact that a constant vertex $U_{sp} $ violates rotational invariance so that one obtains different formulas in the longitudinal and transverse spin channel. The factor  $\tilde{g} =3/8$ comes from averaging longitudinal and transverse channels to restore crossing symmetry (rotational invariance). Here we  explore both values, $\tilde{g} =1/4$ and $\tilde{g} =3/8$, to find out the best agreement with Monte Carlo benchmarks. 

The Migdal-Galitskii relation between single-particle and two-particle properties
%%%%%%%%%
\begin{equation}
    \frac{T}{N}\sum_{\mathbf{k},ik_n} \Sigma_{\sigma}(\mathbf{k},ik_n) \mathcal{G}_{\sigma}(\mathbf{k},ik_n) = U \langle n_{\uparrow}n_{\downarrow} \rangle
		\label{eq:Sigma_G_rule}
\end{equation}
%%%%%%%%
 can be used as an accuracy check ~\cite{Vilk1997}.
In TPSC, this relation is satisfied for the double-occupancy found at the previous level if $\Sigma_\sigma^{(2)}$ from \eref{eq:selfEnergy2} is used with the Green function $\mathcal{G}_{\sigma} $ replaced by $\mathcal{G}^{(1)}_{\sigma} $. 
The difference (generally a few percent) with the calculation where  $\mathcal{G}_{\sigma}^{(2)} $ is used instead, serves as an accuracy check~\cite{Vilk1997}. 
Note that when $\tilde{g} =3/8$, each of the contributions from the three components of the spin fluctuation channels and from the single charge fluctuation channel contribute equally to the Migdal-Galitskii relation. 
Decreasing $\tilde{g}$ gives more importance to the charge contribution to this relation. 

\subsection{Pseudogap and renormalized classical regime in TPSC}
 The self-consistency condition for the spin susceptibility \eref{eq:sumrule_sp} leads to the suppression of the antiferromagnetic mean field phase transition. 
 This is because a finite $T$ phase transition ($\xi=\infty$) and the large phase space occupied by long-wavelength fluctuations in two dimensions would cause a divergence of the integral on the left-hand side of~\eref{eq:sumrule_sp}. 
 Consequently, the phase transition is replaced by a crossover to the renormalized classical regime $\omega_{sp}  < T$ that exists in a wide temperature range. 
 Here the characteristic spin-fluctuation frequency $\omega_{sp} \sim \xi^{-2}$ is small because of critical slowing down. 
 In this regime, the correlation length grows exponentially and quickly overcomes the characteristic thermal wavelength of electrons $ \xi_{th}$.
 The latter increases only as $v_F/T$ for regular points on the Fermi surface \cite{Vilk1996} and as $1/T^{1/2}$ for the Van Hove point $v_F=0$ \cite{vilk2023criteria}. 
 When the correlation length becomes $ \xi >> \xi_{th}$ a pseudogap opens up in the single-particle spectra. 
 This pseudogap is a precursor to the antiferromagnetic gap in the ground state. 
 The regimes described above where studied in detail and benchmarked in Ref.~\cite{Schaefer2021}. 

 Close to the quantum-critical point where antiferromagnetism disappears, the correlation length grows only algebraically with decreasing temperature~\cite{Bergeron:2012} so that it does not overcome quickly the thermal de Broglie wavelength. 
 In that regime, a different criterion for the pseudogap based on second derivative of the spectral weight at $\omega=0$ may be preferable~\cite{Ye_Chubukov_2023}. 

  While TPSC correctly describes the beginning of the renormalized classical regime and opening of the pseudogap, it fails deep in the pseudogap regime. 
  In  that regime, $U_{sp}$ stays just slightly below $U_{crit}=2/\chi^{(1)}(\mathbf{Q},0)$ where $\mathbf{Q}=(\pi,\pi)$. 
  Since $\chi^{(1)}(\mathbf{Q},0)$ diverges as $T \rightarrow 0 $ at half-filling in the near-neighbor hopping model,  $U_{crit} $ and thus $U_{sp}$ and double occupancy (\eref{eq:ansatz_tpsc}) go to zero as well.
  This is, obviously, nonphysical. 
  To remedy this behavior, it was suggested in  Ref.~\cite{gauvin2023improved} to replace $\chi^{(1)}(\mathbf{Q},0)$ with the quantity that remains finite as $T \rightarrow 0 $. We discuss this approach in the next section. 

\subsection{TPSC+ approach}	
 In the TPSC+ approach, the bubble $\chi^{(1)}(\mathbf{q},iq_n) $ in \eref{eq:chispqiqn} and \eref{eq:chichqiqn} is replaced by $\chi^{(2)}(\mathbf{q},iq_n) $, which is calculated as:
\begin{align}
    \chi^{(2)}(\mathbf{q},iq_n&) =  \nonumber \\
    -\frac{T}{N}&\sum_{\mathbf{k},ik_n}\left((\mathcal{G}_{\sigma}^{(2)}(\mathbf{k},ik_n)\mathcal{G}_{\sigma}^{(1)}(\mathbf{k}+\mathbf{q},ik_n+iq_n)\right.\nonumber \\
     &+\left. \mathcal{G}_{\sigma}^{(2)}(\mathbf{k},ik_n)\mathcal{G}_{\sigma}^{(1)}(\mathbf{k}-\mathbf{q},ik_n-iq_n) \right ).
    \label{eq:chi2}
\end{align}
%%%%%%%%%%%%%%%%
 All other self-consistency conditions are the same as in the TPSC and $\mathcal{G}_{\sigma}^{(2)}$ is calculated iteratively using the self-energy \eref{eq:selfEnergy2}. 
 The approach is analogous to the pairing approximation (GG0 theory) for the pair susceptibility introduced by Kadanoff and Martin \cite{Kadanoff_Martin_1961, chen2005bcs, Boyack_2018}. 
 This approach partly takes into account self-energy and vertex corrections~\cite{Schaefer2021,gauvin2023improved}.

 It was shown in Ref.~\cite{gauvin2023improved} that the above form of $\chi^{(2)}(\mathbf{q},iq_n)$ remains finite as $ T \rightarrow 0 $ due to the opening of the pseudogap and that the value of the gap $\Delta$ is given by a modified Stoner criterion with bare $U$ replaced  by the renormalized interaction $U_{sp}$, the irreducible vertex in the spin channel. 
 
 For convenience, we  summarize the main qualitative results at half-filling. The generalized Stoner criterion takes the form:
\begin{equation}
    \frac{2}{U_{sp}} = \int d\epsilon \frac{\rho(\epsilon)}{\sqrt{\epsilon^2+\Delta_0^2}},
    \label{eq:Delta_MF}
\end{equation}
%%%%%%%%%%%%%%%%
where the density of states $\rho(\epsilon)$ is
%%%%%%%%%%%%%%%%
\begin{equation}
    \rho(\epsilon) =\frac{1}{2\pi^2}K\left [ 1-\left (\frac{\epsilon}{4} \right)^2\right] ,
    \label{eq:dos}
\end{equation}
%%%%%%%%%%%%%%%%
with $K$ the complete elliptic integral of the first kind (See \footnote{The complete elleptic integral of the first kind is defined either as $K(m)=\int_0^{\pi/2}[1-m^y sin(t)^2]^{-1/2}dt$ with $y=1$ or $y=2$. The most common definition, including in $\rm{scipy}$, is with $y=1$, which we adopt and which gives the correct value of the density of states on the square lattice. Older literature for that density of states uses the convention $y=2$, which is also correct as long as the corresponding tables are used.} for the ambiguity in definition). 

 Equation (\ref{eq:Delta_MF}) for the Stoner criterion was obtained assuming that the self-energy in the renormalized classical regime is given by 
its leading contribution: the classical thermal fluctuations only. 
The latter corresponds to the zero Matsubara frequency contribution in \eref{eq:selfEnergy2}
%%%%%%%%%%
\begin{align}
    \Sigma_{cl}(\mathbf{k},ik_n) = \frac{T}{N} \tilde{g} U U_{sp} \sum_{\mathbf{q}} \chi_{sp}(\mathbf{q},0) \mathcal{G}_\sigma^{(1)}(\mathbf{k}+\mathbf{q},ik_n).		
    \label{eq:selfEnergy_Cl}
\end{align} 
%%%%%%
In the renormalized classical regime, $\chi_{sp}(\mathbf{q},0) $ is strongly peaked at $\mathbf{q}=\mathbf{Q}$ and in two dimensions, phase space is such that one can replace $\mathcal{G}_\sigma^{(1)}(\mathbf{k}+\mathbf{q},ik_n)$ by $\mathcal{G}_\sigma^{(1)}(\mathbf{k}+\mathbf{Q},ik_n)$. As a result, the self-energy takes the form:
%%%%
\begin{equation} \label{eq:sEnergy_k_Delta}
 \Sigma(\mathbf{k},ik_n)= \frac{\Delta^2(T)}{ik_n +\epsilon(\mathbf{k})},
\end{equation}
%%%%%%
where we took into account that at half-filling $\epsilon(\mathbf{k} +\mathbf{Q})=-\epsilon(\mathbf{k})$ and that $\frac{T}{N} \sum_{\mathbf{q}} \chi_{sp}(\mathbf{q},0) $ is finite in the renormalized classical regime. 
Substituting the self-energy expression \eref{eq:sEnergy_k_Delta} into  \eref{eq:chi2} for $\chi^{(2)}$ and using the zero temperature phase transition condition $U_{sp} \chi^{(2)}(\mathbf{Q},0)/2 =1$, one obtains the generalized Stoner criterion \eref{eq:Delta_MF} for the zero temperature gap $\Delta_0$.  	

\subsection{TPSC+Classical}
 Using the classical contribution to the self-energy \eref{eq:selfEnergy_Cl} to feedback in $\chi^{(2)}$ is sufficient to obtain qualitatively the same results as those obtained in  TPSC+. 
 Moreover,  we find that  the resulting spin susceptibility gives  quantitative results very similar to those obtained using the full self-energy \eref{eq:selfEnergy2} (see next section). 
 This suggests some cancellation between vertex and quantum self-energy corrections. 
 We  call this version of the TPSC+ approach, TPSC+Classical. 
 It uses minimal necessary feedback self-energy to get good description of the spin susceptibility deep in the renormalized classical regime. The use of momentum and frequency dependent feedback self-energy together with constant vertex involves trade off. It gives better description of the antiferromagnetic fluctuation but it is not consistent with conservation laws. It thus make sense to keep the feedback self-energy in the calculation of the susceptibility to a minimum. 

  We stress that the final self-energy obtained in the TPSC+Classical approach is still the one that includes both spin and charge quantum fluctuations in the self-energy, as defined in \eref{eq:selfEnergy2}. 
  Only the feedback self-energy used in the calculation of the partially-dressed susceptibility $\chi^{(2)}$ takes the form defined in \eref{eq:selfEnergy_Cl}. 

 In the TPSC+Classical approach the Migdal-Galitskii sum rule \eref{eq:Sigma_G_rule} is not satisfied exactly, but the violation is only a few percent even deep in the renormalized classical regime.
 The f-sum rule for the susceptibility is violated by  TPSC+Classical. 
 In particular, $\chi^{(2)}(0,iq_n) \neq 0 $ for  $q_n \neq 0$ which leads to non-zero values of $\chi_{sp}(0,iq_n) $ and $\chi_{ch}(0,iq_n) $ for  $n \neq 0$. 
 The ratio $\chi^{(2)}(0,iq_1)/\chi^{(2)}(0,iq_0)$ is less than $10\%$ above  the renormalized classical regime and less than $30\%$ in  that regime.  
 Despite these drawbacks, the results in the next sections show some of the virtues of this approach.

\section{Results} 
\label{sec:results}
Thanks to improvements in the numerical algorithms, explained in Appendix \ref{app:Num_Impl} we can compare our results with DiagMC as a benchmark, in regimes that were previously inaccessible to TPSC+. 
Analytic continuation then allows us to exhibit the spectral weight deep in the renormalized classical regime. 

\subsection{Comparison with DiagMC results} 
\label{sec:Comp_with DiagMC}

\subsubsection{Spin Susceptibility}

\begin{figure}
    \centering
    \includegraphics[width=\columnwidth]{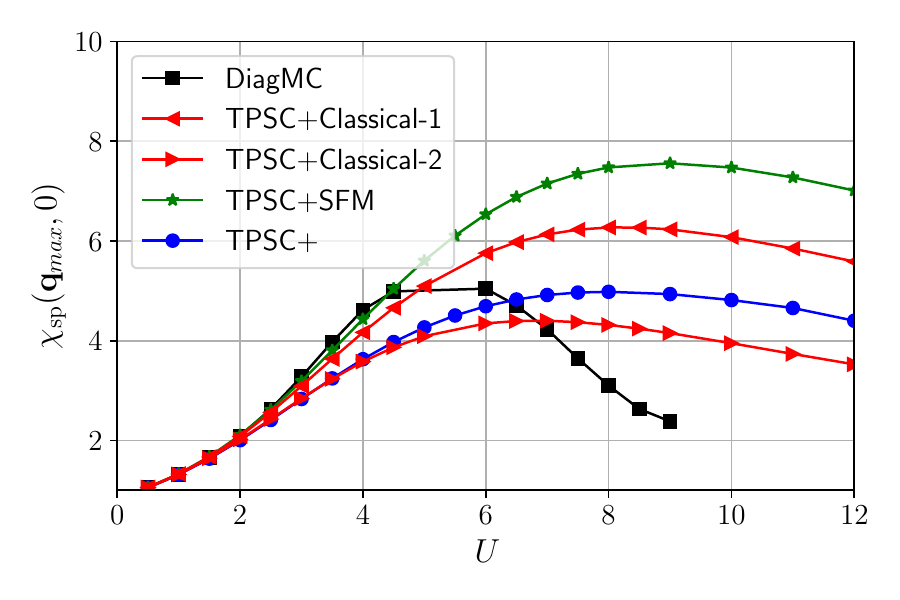} 
     \caption{Maximal value of the spin susceptibility obtained from various
TPSC+ approaches (see text) as a function of $U$ compared with Monte Carlo data \cite{iv_two-dimensional_2022}. Results are shown for  $n = 0.875$ and $T = 0.2$. All TPSC+ approaches agree reasonably well with the benchmark Monte Carlo data for $U <=2.5$ at this temperature. For $ U > 4 $ the Monte Carlo values of $\chi_{sp}(\mathbf{q}_{max}, 0)$ start to saturate and then decrease as a function of $U$. The latter behavior indicates transition to Mott physics. All TPSC+ approaches miss this behavior quantitatively but they do show a broad maximum for $U \geq 6$. This maximum is due to the effect of the feedback self-energy on the spin susceptibility.}
    \label{fig:ch_sp_n_0_0875_U_T_0_2}
\end{figure}

We start with results for the maximum of spin susceptibility as a function of the interaction strength.  \fref{fig:ch_sp_n_0_0875_U_T_0_2} shows $\chi_{sp}(\mathbf{q}_{max}, 0)$ as a function of $U$ for $n=0.875$ and $T=0.2$. In this figure TPSC+Classical-1 uses $\tilde{g}_{fb}=1/4$ for feedback self-energy and TPSC+Classical-2 uses $\tilde{g}_{fb}=3/8$. The TPSC+ version uses full self-energy \eref{eq:selfEnergy2} with both spin and charge contributions and $\tilde{g}_{fb}=3/8$ while TPSC+SFM uses only spin contribution with $\tilde{g}_{fb}=1/4$. As we can see from figure \fref{fig:ch_sp_n_0_0875_U_T_0_2}, TPSC+Classical-1 has almost the same results as TPSC+SFM while TPSC+Classical-2 has almost the same results as TPSC+ for $ U \leq 4 $. This implies that quantum contributions to the feedback self-energy has rather small effect on the spin susceptibility  $\chi_{sp}(\mathbf{q}_{max}, 0)$ for these values of  $U $. We found similar results for other fillings $n$ and temperatures $T$ and $ U \leq 4 $. This suggests some cancellation between self-energy and vertex corrections. 
The area of the applicability for all TPSC+ variants is weak to intermediate interaction regime $ U \leq 4 $. Comparison with Monte Carlo results \cite{iv_two-dimensional_2022} in the \fref{fig:ch_sp_n_0_0875_U_T_0_2} shows that for $ U > 4 $ the Monte Carlo values of $\chi_{sp}(\mathbf{q}_{max}, 0)$ start to saturate and then decline as a function of $U$. 
The latter behavior indicates transition to the Mott physics (localized spins) in the strong coupling limit. All TPSC+ variants miss this behavior quantitatively. 
We note, however, that all TPSC+ variants do have a maximum of $\chi_{sp}(\mathbf{q}_{max}, 0)$ as a function of $U$ but it occurs at larger $ U \geq 6 $ and it is much broader than Monte Carlo result. Nevertheless, it is interesting because in the original TPSC theory $\chi_{sp}(\mathbf{q}_{max}, 0)$ saturates with increasing $ U $ but does not decrease. 
Thus the decrease of $\chi_{sp}(\mathbf{q}_{max}, 0)$ at large $U$ in our theory is solely due to the effect of the feedback self-energy on the spin susceptibility. 
In the opposite limit of relatively small $ U \leq 2.5 $, all TPSC+ variants agree quantitatively with Monte Carlo data.    

The TPSC+ study in the Ref.~\cite{gauvin2023improved} has results for various fillings, temperatures and interaction strength for both two and one particle properties. The results are for TPSC+ and TPSC+SFM but, as we mentioned above, the results for $ U \leq 4 $ are quantitatively similar to results for TPSC+Classical with $\tilde{g}_{fb}=3/8$ and $\tilde{g}_{fb}=1/4$, respectively. For technical reasons, study in the Ref.~\cite{gauvin2023improved} was limited to temperatures $T \geq 0.79$.  
In the current study, we were able to resolve convergence problem at low temperatures using methods that are explained in Appendix~\ref{app:Num_Impl}. 
In the rest of the paper, we focus on low temperatures at half-filling, on the antiferromagnetic pseudogap regime and on estimates of the zero-temperature gap. 
We choose $U=2$ because this allows us to make quantitative comparison to the DiagMC results. 
We also found that at very low temperatures (large correlation length) the TPSC+Classical with $\tilde{g}_{fb}=3/8$ works the best and we use this value in the rest of the paper.

 \subsubsection{Double Occupancy}
\begin{figure}
    \centering
    \includegraphics[width=\columnwidth]{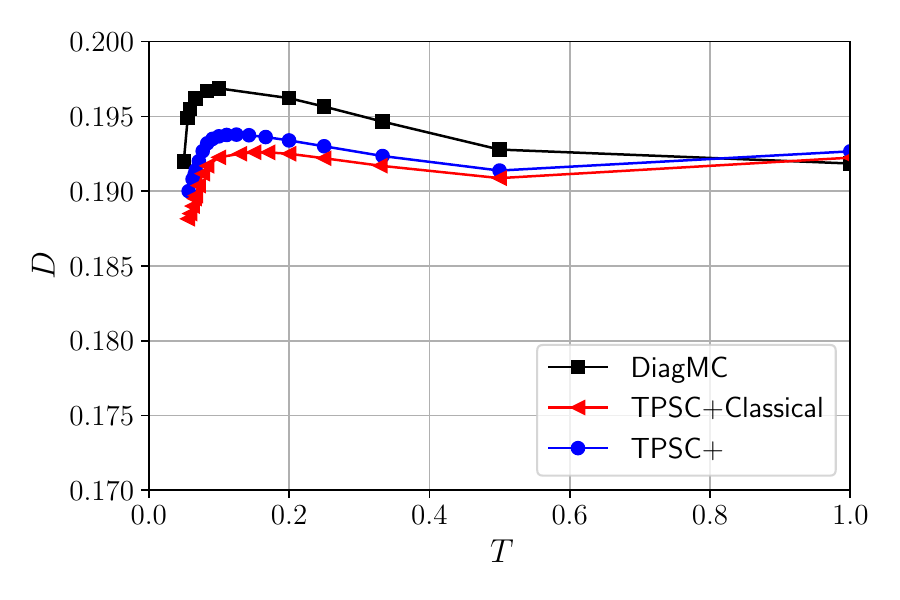} 
    \caption{Double occupancy as a function of the temperature $T$ for $n=1$ and $U=2$. Results for both TPSC+ and TPSC+Classical approaches are in good qualitative and quantitative agreement with DiagMC \cite{Schaefer2021} results. Importantly, they capture a small rise and then drop of the double occupancy as $T$ enters the renormalized-classical  regime}
    \label{fig:doccup}
\end{figure}

  Figure \ref{fig:doccup} shows  double occupancy $D$ as a function of  temperature $T$. The results are for   TPSC+, TPSC+Classical and for the DiagMC \cite{Schaefer2021} benchmark results. The factor  $\tilde{g}_{fb}$ for the feedback self-energy used in the calculation of the partially-dressed susceptibility $\chi^{(2)}$ was set to $\tilde{g}_{fb} =3/8$. Both TPSC+ approaches perform well qualitatively and even quantitatively within about $2\%$ . In particular, they capture a small low-temperature drop of double occupancy in the renormalized classical regime. 
	
\subsubsection{Correlation Length}
\begin{figure}
    \centering
    \includegraphics[width=\columnwidth]{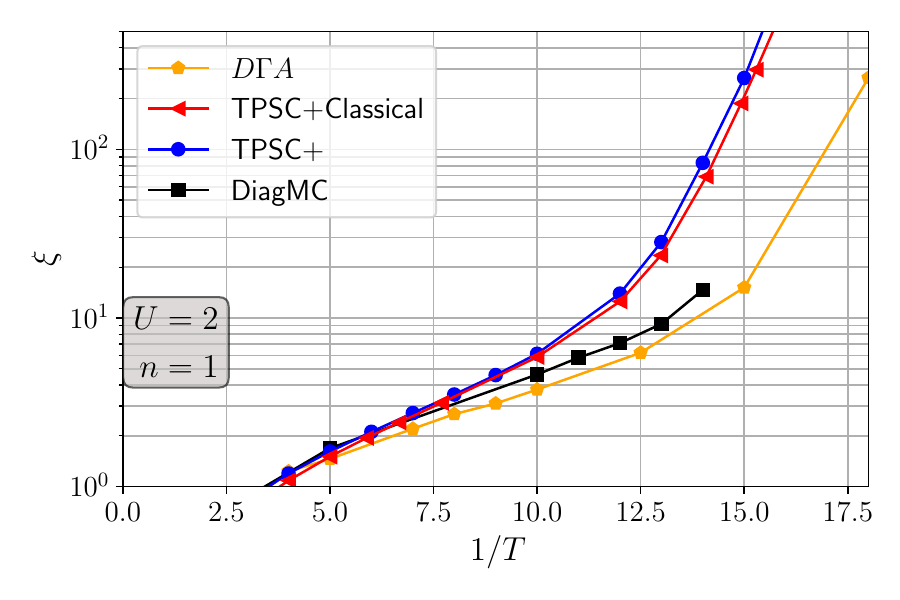} 
    \caption{Spin correlation length from TPSC+ and TPSC+Classical approaches compared
with the DiagMC benchmark and $D\Gamma A$ approximation. The results are obtained as a function of the inverse temperature $\beta=1/T$ for the half-filled 2D Hubbard model with U = 2.}
    \label{fig:xi_beta}
\end{figure}

%%%%%%%%%%%
Figure \ref{fig:xi_beta} shows the spin correlation length $\xi(\beta=1/T)$ defined by ~\cite{gauvin2023improved}
%%%%%%%%%%%%%%
 \begin{equation} \label{eq:xi_sp}
    \xi = \xi_0\sqrt{\frac{U_{sp}}{\frac{2}{\chi^{(2)}(\mathbf{Q},0)}-U_{sp}}},
\end{equation}
%%%%%%%%%%%%%
where the bare particle-hole correlation length is 
%%%%%%%%%%%%%%%%
\begin{equation} \label{eq:xi_0}
    \xi_0^2 = \frac{-1}{2\chi^{(2)}(\mathbf{Q}, 0)}\left .\frac{\partial^2\chi^{(2)}(\mathbf{q},0)}{\partial q_x^2}\right|_{\mathbf{q}=\mathbf{Q}}.
\end{equation}
In addition to the  TPSC+, TPSC+Classical approaches and DiagMC results, there are also results for the $D\Gamma A$ approximation \cite{Schaefer2021}. The $D \Gamma A$ approximation has data for lower temperatures than DiagMC. Both TPSC+ approaches are in good agreement with the DiagMC and the $D \Gamma A$ approaches at high temperatures (small $\beta$). As temperature decrease, the correlation length enters an exponential regime. The onset of this regime $ T_{exp}, $ occurs at higher temperature for TPSC+ than for $D \Gamma A$. The lowest $ T_{exp} $ is for  TPSC+Classical but the difference with  TPSC+ is small. 

\subsubsection{Self Energy}
\begin{figure}
    \centering
    \includegraphics[width=\columnwidth]{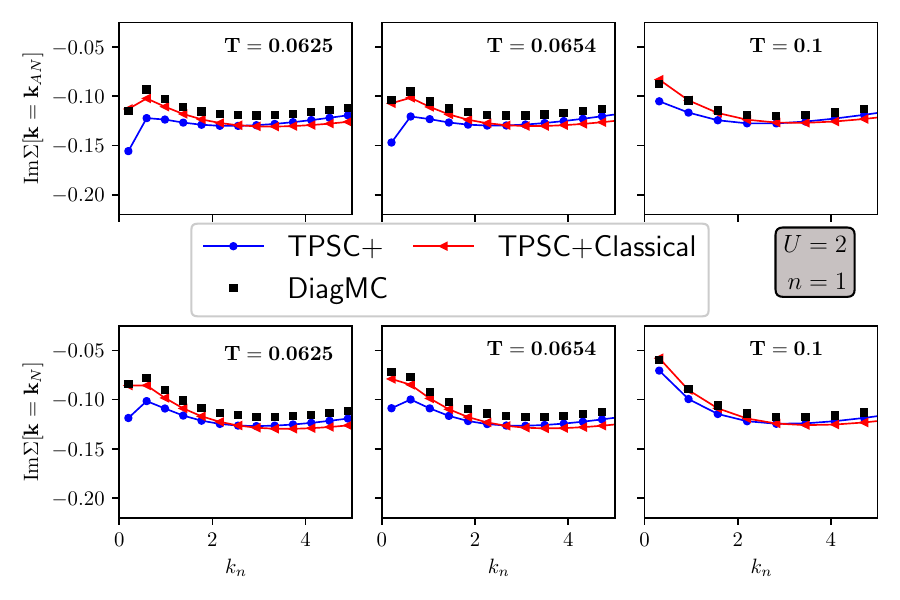} 
    \caption{The imaginary part of the self-energy as a function
    of  Matsubara frequencies at half-filling $n = 1$, and interaction $U = 2$ for three temperatures $T=0.0625, 0.0654, 0.1$ and nodal (N) $\mathbf{k}_{N}=(\pi/2,\pi/2)$ and antinodal (AN) $\mathbf{k}_{AN}=(\pi,0)$ points on the Fermi surface. 
    The symbols are DiagMC results \cite{Schaefer2021}. 
    The drop of the self-energy at the smallest Matsubara frequency is due to the pseudogap.}
    \label{fig:SE_omega}
\end{figure}

   Figure \ref{fig:SE_omega} shows the imaginary part of the self-energy as a function of  Matsubara frequency at half-filling $n = 1$, and interaction $U = 2$ for three temperatures $T=0.0625, 0.0654, 0.1$ and nodal (N) $\mathbf{k}_{N}=(\pi/2,\pi/2)$ and antinodal (AN) $\mathbf{k}_{AN}=(\pi,0)$ points on the Fermi surface. 
   The self-energy was calculated using the full expression for the second level approximation \eref{eq:selfEnergy2} with $\tilde{g}=1/4$ for TPSC+Classical and $\tilde{g}=3/8$ for TPSC+. 
   The drop of the self-energy at the smallest Matsubara frequency is due to the pseudogap. This was used as a criterion for the pseudogap in Ref.~\cite{Schaefer2021}. 
   
   According to the DiagMC results \cite{Schaefer2021} in Matsubara frequency, the crossover temperatures to the pseudogap regime (pseudogap) are $T=0.0654$ for the AN point and $T=0.0625$ for the N point. 
   However, as we see in the next section, the real-frequency results show a pseudogap at higher temperatures {\it i.e.} before $|\Sigma(k_F, ik_0)| $ becomes larger than $|\Sigma(k_F, ik_1)| $. 
   The TPSC+Classical results with $\tilde{g}=1/4$ are in quantitative agreement with DiagMC results \cite{Schaefer2021}. 
   The TPSC+, which uses $\tilde{g}=3/8$, captures the pseudogap effect qualitatively but overestimates it quantitatively. 
   In the rest of the paper we display results for TPSC+Classical with $\tilde{g}=1/4$. 

\begin{figure}
    \centering
    \includegraphics[width=\columnwidth]{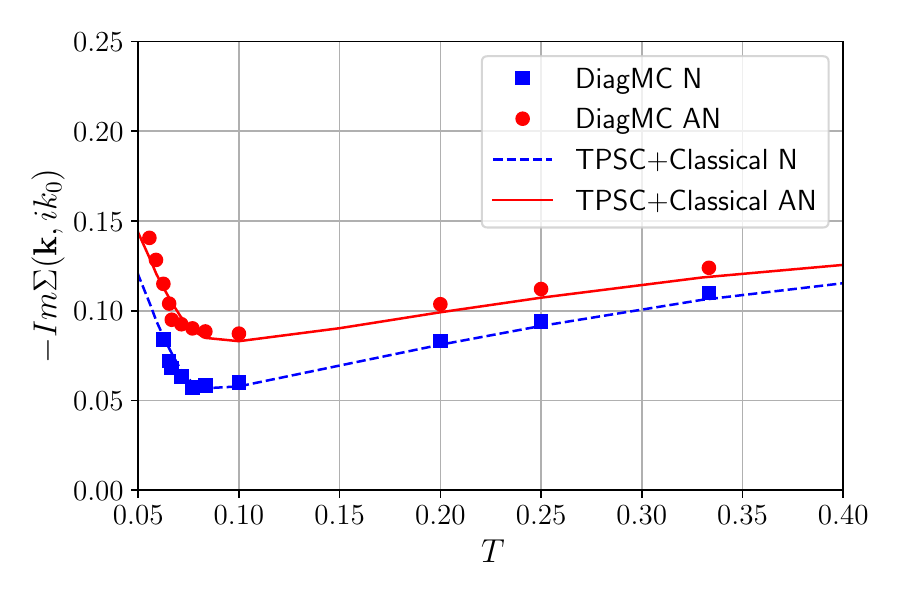} 
    \caption{The negativeimaginary part of the self-energy, for $ U=2$, $n=1$, at the first Matsubara frequency for antinode and node as a function of temperature for the numerically exact DiagMC (symbols) and the TPSC+Classical (lines). 
    The TPSC+Classical results are in very good agreement with DiagMC results in both the metallic and the the pseudogap regimes. 
    The rapid increase of this quantity with decreasing temperature is strong evidence of the pseudogap. 
    The analytic theory predicts divergence of the self-energy  as $ T \rightarrow 0 $. }
    \label{fig:SE_0_T}
\end{figure}

Figure \ref{fig:SE_0_T} shows the absolute value of the self-energy at the lowest Matsubara frequency  $-\mathrm{Im} \Sigma(\mathbf{k},i k_0)$  as a function of temperature for the nodal and antinodal points on the Fermi surface. 
The symbols are data from DiagMC \cite{Schaefer2021} and the lines represent results of the TPSC+Classical calculations.  
Agreement is very good. 
TPSC+Classical captures quantitatively the behavior in both the metallic ($T$ above minimum) and the pseudogap regimes ($T$ below minimum).
%%%%%%%%%%%%%
%%%%%%%%%%%
  Figure ~\ref{fig:SE_0_T} presents the clearest evidence of the pseudogap in the Matsubara representation because this quantity is expected to diverge as $T \rightarrow 0$ (see \eref{eq:sEnergy_k_1} below). 
  Indeed, the figure shows this tendency as temperature decreases. 
  Note also that the absolute value of the self-energy $\mathrm{Im} \Sigma(\mathbf{k},i k_0)$ is larger for the antinodal point than for the nodal point. 
  For the pseudogap regime this is expected from the asymptotic behavior of $\mathrm{Im} \Sigma(\mathbf{k},i k_0)$ for the antinodal and the nodal points. 
  The asymptotic behavior of $\mathrm{Im} \Sigma(\mathbf{k},i k_0)$ is given by the classical contribution to the self-energy \cite{Vilk1997}, \cite{vilk2023criteria}: 
  For the antinodal point:
%%%%%%%%%%%%%%		
	\begin{equation} \label{eq:sEnergy_k_1}
 \Sigma_{cl}(\mathbf{k}_{AN},ik_0)= \frac{\Delta^2}{i\pi T}\left[ 1+ \frac{T}{2T_0} \ln \left( \frac{2 \pi T}{w \tilde{\xi}_0^{-2}} \right)\right],
\end{equation}
where $T_0$ is the characteristic temperature for the exponentially growing regime, the spin correlation length $\xi$ is $\xi=\tilde{\xi}_0 \exp(T_0/T)$, the gap parameter  $\Delta$ is $\Delta^2=\frac{\tilde{g} U T_0}{ \pi \xi_{0}^{2} } $ and $w=(1/2)\partial^2\tilde{\epsilon}(\mathbf{k}+\mathbf{Q})/\partial k_x^2 $. 
%%%%%%%%%%%%%%%%%%%%	
For the nodal point:
%%%%%%%%%%%%%%%%%%%%%%%	
\begin{equation}
    \Sigma_{cl}(\mathbf{k}_{N},ik_0)=\frac{\Delta^2}{i\pi T}\left[ 1+ \frac{T}{T_0} \ln \left( \frac{2\pi T}{v_F \tilde{\xi}_0} \right)\right].
    \label{eq:sEnergy_st_R_M_Delta}
\end{equation}
%%%%%%%%%%%%%%%%%%
  One can see from the expressions \eref{eq:sEnergy_k_1}, \eref{eq:sEnergy_st_R_M_Delta} that the leading term is the same for the AN and the N points but the main temperature correction $ \sim T \ln T$ is smaller by a factor of two for the AN point than for the N point. 
  %%%%%%%%%%%%%%%%%%%%%%%%%%
\subsubsection{Zero-Temperature Gap}
\begin{figure}
    \centering
    \includegraphics[width=\columnwidth]{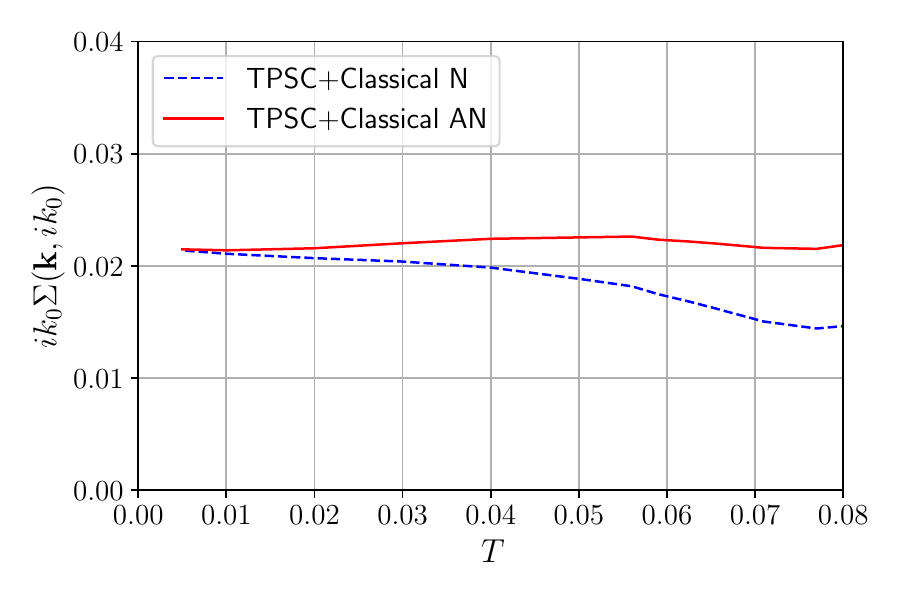} 
    \caption{The quantity $ ik_0 \Sigma(\mathbf{k},i k_0)$ as a function of the temperature, for $U=2, n=1$. As expected from the analytical results, the curves for the nodal and the antinodal points plateau and converge at low temperature.}
    \label{fig:Delta_sq_T}
\end{figure}

 The above expressions suggest a way to find the zero-temperature gap $\Delta_0$. 
 We plot the quantity $ ik_0 \Sigma(\mathbf{k},i k_0)$ for the nodal and the antinodal points as a function of  temperature. 
 The temperature where both lines converge to the same asymptote is the temperature at which we can extract the value of $\Delta_0^2=ik_0 \Sigma(\mathbf{k},i k_0)$. 
 Figure \ref{fig:Delta_sq_T} illustrates the above approach. 
 The lines for the nodal and antinodal points converge as temperature decreases when $T$ reaches $T\sim 0.005$. 
 From this we obtained the value  $\Delta_0 \approx 0.15 $.
%%%%%%%%%%%%%%%%%%%%%%	
%%%%%%%%%%%%%

It is interesting to compare this result with the renormalized mean-field one, obtained from the generalized Stoner criterion \eref{eq:Delta_MF}. 
Solving \eref{eq:sEnergy_k_Delta}, we obtain $\Delta_{0\_MF} \approx 0.18$. 
This is about $20\%$ larger than the result $\Delta_0 \approx 0.15$ obtained using the above procedure ($\Delta_0^2=ik_0 \Sigma(\mathbf{k},i k_0)$ as $T \rightarrow 0$).
Our analysis indicates that this difference is mostly because we used different values for factors  $\tilde{g}_{fb}=3/8$ and $\tilde{g}=1/4$.
Choosing the same value for both factors  $\tilde{g}_{fb}=3/8$ and $\tilde{g}=3/8$ gives the gap value within $1\%$ of the mean field result $\Delta_{0\_MF} \approx 0.18$.
We used the combination of factors $\tilde{g}_{fb}=3/8$ and $\tilde{g}=1/4$ because it gives the best fit of the Monte Carlo data for both spin susceptibility and the final self-energy. 
Since we have a good fit of the self-energy $\Sigma(\mathbf{k},i k_0)$ to the Monte Carlo data \ref{fig:SE_0_T}, we believe that the value of $ \Delta_0=0.15 $ extracted from the above self-energy analysis is more accurate than the renormalized mean-field result. The mean-field gap \eref{eq:Delta_MF} was derived using classical critical fluctuations only and the deviation from it gives a measure of quantum effects.

%%%%%%%%%%%%%%%%%%
\subsection{Analytic continuation to real frequencies} 
\label{sec:Analytical_continuation}

 In this section, we present results of analytic continuation from the Matsubara frequency representation to real frequencies obtained from the Padé approximation. 
 The interaction strength and filling are the same as in the previous section: $U=2$ and $n=1$.
 
\begin{figure}
    \centering
    \includegraphics[width=\columnwidth]{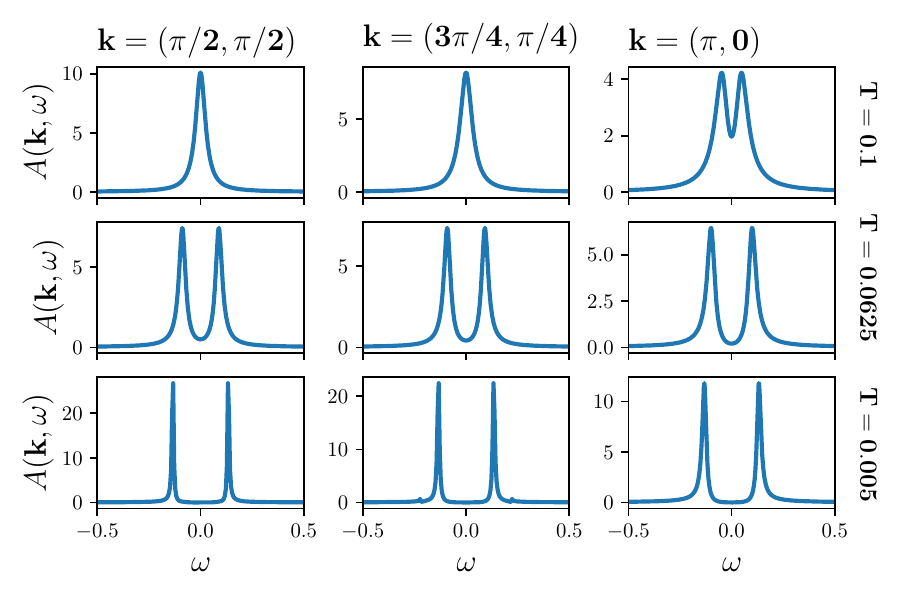} 
    \caption{Evolution of the spectral function $A(\mathbf{k}, \omega) $ with temperature. The results are for tree points on the Fermi surface. 
    The parameters are $U=2$ and $n=1$. The upper panel shows that at the antinodal point $\mathbf{k}_{AN}=(\pi,0)$ ($v_F=0$) the pseudogap has started to develop for $T=0.1$. 
    This is significantly higher than the pseudogap crossover temperature identified earlier \cite{Schaefer2021} based on  Matsubara data only. 
    For regular points on the Fermi surface ($v_F \neq 0$) there is still a single particle peak in $A(\mathbf{k}, \omega) $ at this temperature. 
    The middle panel $T=0.0625$ shows the pseudogap in the spectral function at all points on the Fermi surface. 
    The lower panel shows a well developed pseudogap at the very low temperature $T=0.005 < \Delta $ that is hard, visually, to distinguish from a real antiferromagnetic gap.}
    \label{fig:Akw}
\end{figure}

 Figure \ref{fig:Akw} shows the evolution of the spectral function $A(\mathbf{k}_F, \omega) $ with temperature. 
 The results are presented for three points on the Fermi surface: the nodal point $\mathbf{k}_N=(\pi/2,\pi/2)$, the antinodal (Van Hove) point $\mathbf{k}_{AN}=(\pi,0)$ and  the point in between these points $\mathbf{k}=(3 \pi/4,\pi/4)$. 
 
 At temperature $T=0.1$ (upper panel) a single quasi-particle peak in the spectral function $A(\mathbf{k}, \omega)$ exists for regular points on the Fermi surface ($v_F \neq 0 $), but for the Van Hove point $\mathbf{k}_{AN}=(\pi,0)$ the pseudogap is already opening up. 
 Note that this is significantly higher than the crossover temperature $T=0.0654$ identified in Ref.~\cite{Schaefer2021} based on Matsubara frequency data only. 
 The criterion for the pseudogap used in \cite{Schaefer2021} was $|\Sigma(\mathbf{k}_F, ik_0)| > |\Sigma(\mathbf{k}_F, ik_1)| $. 
 However, the real frequency data provides a different perspective and we conclude that the condition $|\Sigma(\mathbf{k}_F, ik_0)| > |\Sigma(\mathbf{k}_F, ik_1)| $ is a sufficient but not a necessary condition for the pseudogap.
 In the case we study, this criterion gives a lower bound.
 We also verified that for the nodal point the crossover to the pseudogap in $A(\mathbf{k}_N, \omega) $ occurs at the temperature $T_N=0.08$. 
 This is also higher than the crossover $T_N$ identified based on the above Matsubara-frequency criterion ($T_N=0.0625$). 
 The difference between crossover temperatures for antinodal point $T_{AN}$ and  nodal point $T_{N}$ is about $ (T_{AN} - T_N) \approx 0.02$, which is much larger than the difference based on the Matsubara criterion  $ (T_{AN} - T_N) \approx 0.003  $. 
 
 At temperature $T=0.0625$ (middle panel), the pseudogap is present at all points on the Fermi surface and it can easily be distinguished from an antiferromagnetic gap. 
 
 Finally, at $T=0.005$ (bottom panel) the pseudogap is hard to distinguish, visually, from  a real antiferromagnetic gap in the spectral function. 
 The results for $T=0.005 < \Delta $ give insight in the behavior  at $ T \rightarrow 0 $. 
 In addition, to quasiparticle peaks at $\omega = \pm \Delta $ and the pseudogap between them, there is also an incoherent background for $|\omega | > \Delta $. 
 This is consistent with the results for the half-filled Hubbard model obtained using the nonlinear sigma ($NL\sigma M$) description of the spin fluctuations \cite{borejsza2003antiferromagnetism}. 
 The latter work also shows that the peak at $|\omega | \rightarrow \Delta -0 $ becomes infinitely narrow and infinitely high as $ T \rightarrow 0 $. On the other hand, the height of $A(\mathbf{k}_F, \omega) $ remains finite when $|\omega | \rightarrow \Delta +0 $. 
 This behavior implies that the imaginary part of the self-energy $\Sigma''(\mathbf{k}_F, \omega)$ in \cite{borejsza2003antiferromagnetism} is a step function around $|\omega | = \Delta $ with zero value inside the region  $ 0 <|\omega | < \Delta $ and a finite value outside the gap (at exactly $\omega =0 $ there is negative delta function for $ T \rightarrow 0 $)  . 
 A different perspective on the behavior of the self-energy in the ordered state is provided in Ref. \cite{sangiovanni2006static} in which the self-energy is continuous at $|\omega | = \Delta $ and goes to zero as $ T \rightarrow 0 $.  
 The results in Ref. \cite{sangiovanni2006static} are presented for the strong coupling limit and were obtained using dynamical mean-field theory (DMFT) in  infinite dimension.
 
 Our result for $|\omega | < \Delta $ and $ T \rightarrow 0 $ is somewhat different in this respect from the results in \cite{borejsza2003antiferromagnetism} and \cite{sangiovanni2006static}. Specifically, the weight of $A(\mathbf{k}_F, \omega) $ for $|\omega | < \Delta $ is very small but it is not zero. 
 This is because in the one-loop like approximation that we use (\eref{eq:selfEnergy2}, the quantum contribution to $\Sigma''(\mathbf{k}_F, \omega)$ is continuous and is a linear function at $|\omega | >> T $ (see bellow). Although it is small, it does not go to zero as $ T \rightarrow 0 $ . 
 
\begin{figure}
    \centering
    \includegraphics[width=\columnwidth]{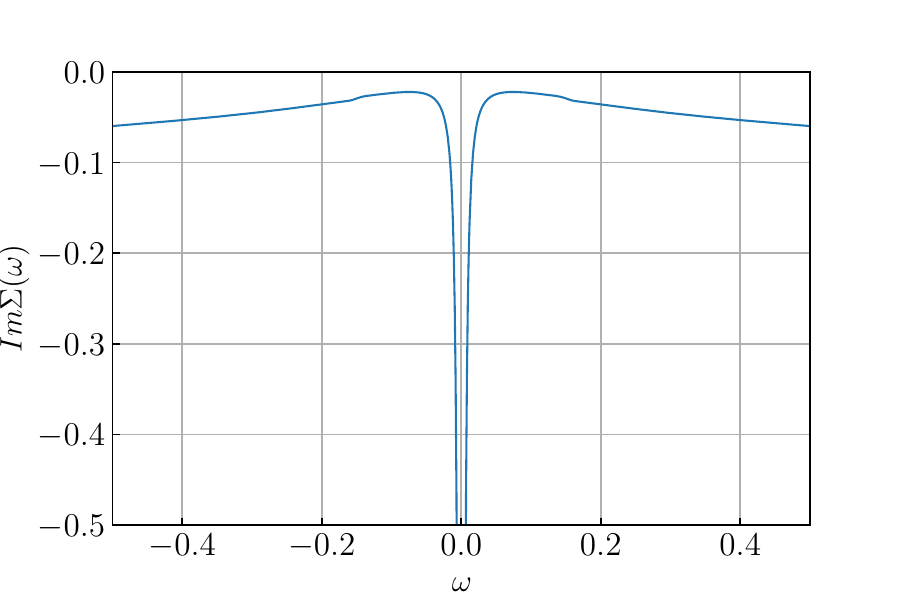} 
    \caption{Imaginary part of the self-energy at the antinode $\Sigma(\mathbf{k}_{AN}, \omega) $ as function real frequency at $T=0.005 $, $U=2$ and $n=1$. 
    The negative delta function like peak at $ \omega=0 $ is due to classical critical fluctuations and is the signature of the pseudogap. 
    The linear frequency dependence for $ | \omega | > \pi T$ is due to quantum fluctuations and nesting. 
    It is finite and continuous at $ |\omega |=\Delta \approx 0.15$.}
    \label{fig:ImSigma_real_omega}
\end{figure}

 Figure \ref{fig:ImSigma_real_omega} displays $\Sigma''(\mathbf{k}_F=\mathbf{k}_{AN}, \omega)$ in our theory for $T=0.005 $ at the antinode. 
 There is a delta-function-like peak  at $ \omega =0 $ (most of the peak is outside the picture) and a linear dependence of $\Sigma''(\mathbf{k}_F=\mathbf{k}_{AN}, \omega)$ for $\omega > \pi T $. 
 The former is due to the critical classical fluctuations and is responsible for the pseudogap. 
 The latter is due to quantum fluctuations. 
 It is linear in $\omega $ due to nesting, it is unrelated to the antiferromagnetic peak in the spin susceptibility and  is qualitatively similar to the second order perturbation theory result~\cite{Lemay2000,Schaefer2021}. 
 We checked explicitly that this quantum linear behavior of $\Sigma''(\mathbf{k}_F, \omega)$ for $ \omega > \pi T $ is not significantly affected by temperature. 
 Since $\Sigma''(\mathbf{k}_F, \omega)$ is finite at $ |\omega| = \Delta \approx 0.15 $ the peaks at $|\omega | \rightarrow \Delta -0 $ remains finite as $ T \rightarrow 0 $. 
 We think that this behavior is likely an artifact of the one-loop-like approximation that we used. 
 The alternative is that the spectral function is not continuous at $ T=0 $ and that the weight inside the gap disappears suddenly at the $ T=0 $ phase transition.     

\begin{figure}
    \centering
    \includegraphics[width=\columnwidth]{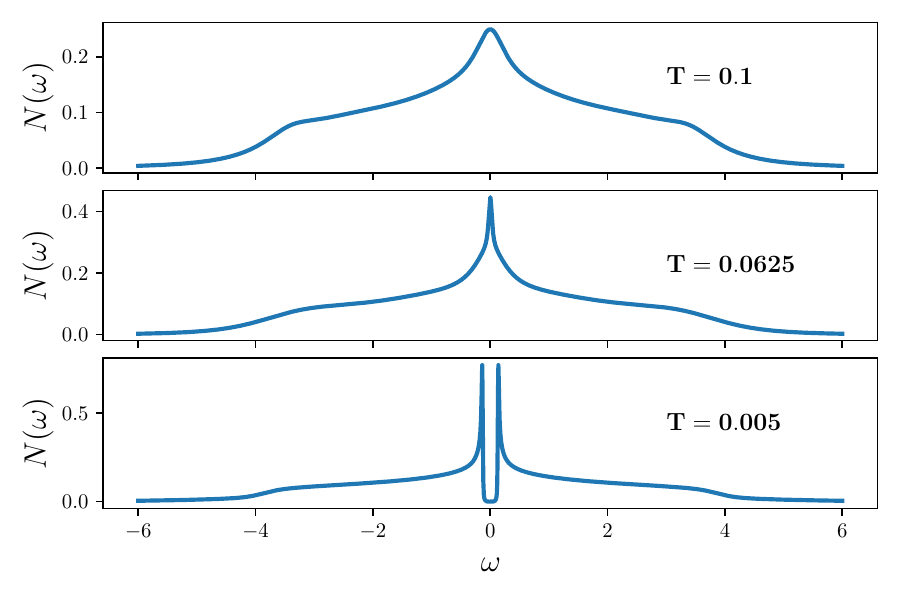} 
    \caption{Evolution of the density of states $ N(\omega) $ with temperature for $U=2, n=1$. 
    The maximum 
		 at $\omega =0 $ at higher temperatures is the remnant of the divergence in $ N(\omega) $ 
		in the non-interacting system (the Van Hove singularity in 2D systems). 
  The pseudogap in 
		the density of states develops at lower temperatures than in the spectral 
		function $A(\mathbf{k}_F, \omega) $ on the Fermi surface.}
    \label{fig:DS_omega}
\end{figure}

 Figure \ref{fig:DS_omega} shows the evolution of the density of states $ N(\omega) $ with temperature. 
 We can see that at higher temperatures there a is a maximum at $ \omega=0$. 
 This maximum is due to the Van Hove singularity that in 2D leads to the divergence of $ N(\omega) $ at $ \omega=0$ in the noninteracting system. 
 Note that at $T=0.0625 $ the pseudogap is already developed in the spectral function $A(\mathbf{k}_F, \omega) $ on the Fermi surface but it is not present in the density of states. 
 This is because that quantity is an integral over all wave vectors, so until the pseudogap is well developed, its Fermi-surface contribution to the total density of states may be negligible. 
 The pseudogap in the density of states eventually develops at lower temperatures and can be clearly seen in the lower panel ($T=0.005$) of \fref{fig:DS_omega}. 
\section{Conclusion}
\label{sec:conclusion}
In this work, we applied the TPSC+ approach for the one-band Hubbard model to the study of the pseudogap induced by antiferromagnetic fluctuations deep in the renormalized-classical regime. 
In addition to  TPSC+, we suggested a slightly modified TPSC+Classical approach. 
Both approaches give very similar results for double occupancy and spin susceptibility despite the fact that TPSC+Classical uses only thermal classical contributions for the feedback of the self-energy into spin fluctuations. 
This suggests some cancellation between vertex and quantum self-energy corrections.

The results for two-particle properties are in semiquantitative agreement with DiagMC and $D \Gamma A $ results in the region where the latter converge. 
We were able to go to much lower temperatures than most other theoretical approaches and to calculate the spin correlation length $\xi$ and spin-susceptibility $\chi_{sp} $ in the thermodynamic limit for temperatures with extremely large correlation lenghts $\xi \leq 10^{6}$. 
This requires special numerical approaches, discussed in Appendix~\ref{app:Num_Impl}.

The computational cost of TPSC+ approaches is low. Even at very low temperatures, it takes about a minute to obtain converged results on a laptop.
Our theory is applicable only in the weak to intermediate interaction regime and it does not satisfy the f-sum rule. 
The later limitation does not seem to affect much the susceptibilities at low frequencies. 
There is an ambiguity in the choice of numerical factors in the feedback $\tilde{g}_{fb} $ and the final $\tilde{g} $ self-energy.  We choose combination of factors $\tilde{g}_{fb} =3/8$ and $\tilde{g}=1/4 $ that gives the best fit to the Monte Carlo data for both the spin susceptibility and the self-energy.

Our three main conclusions are that (a) For the single-particle properties, the TPSC+Classical approach allows us to describe quantitatively DiagMC data in both the metallic and the pseudogap regime. 
(b) The self-energy results at very low temperatures lead to estimates for the zero-temperature gap that is smaller than the renormalized mean-field result (for $U=2, n=1$) by about 20\%, due to quantum fluctuations. 
(c) Analytic continuation of the Matsubara frequency results for the single particle properties using Padé approximants show that the pseudogap in the spectral function opens up significantly earlier than what was determined previously using Matsubara frequency data only and that  the criterion $|\Sigma(\mathbf{k}_F, ik_0)| > |\Sigma(\mathbf{k}_F, ik_1)| $ is a sufficient but not a necessary condition for the pseudogap. 
  
The main puzzle that remains to be solved is the following. In the pseudogap regime, the spectral function at $ T \rightarrow 0 $ has two peaks at $ \omega \pm \Delta $ and an incoherent background due to quantum fluctuations. 
For $ | \omega  | > \Delta $ the presence of the incoherent background is consistent with previous results from the $NL \sigma M$ (nonlinear sigma model) of the antiferromagnetic fluctuations \cite{borejsza2003antiferromagnetism}. 
Our theory also predicts a very small but finite spectral weight for $ |\omega | < \Delta $ and $ T \rightarrow 0 $ due to quantum fluctuations. 
This implies a discontinuity between the zero-temperature limit of the spectral weight in the pseudogap regime and the spectral weight in the zero-temperature long-range antiferromagnetic state. 
The small spectral weight in the antiferromagnetic pseudogap is likely an artifact of the one loop approximation for the self-energy. 
Designing a theory that completely bridges the pseudogap regime and the zero-temperature antiferromagnetic state is an open problem.

\section*{Acknowlegments}
We would like to thank Moïse Rousseau for technical help, Chloé Gauvin-Ndiaye for discussions and for sharing her TPSC code, and David Sénéchal for discussions. 
We acknowledge support by the Natural Sciences and Engineering Council of Canada (NSERC) for funding through RGPIN-2019-05312, and by the Canada First Research Excellence Fund. 

\appendix

\section{Numerical implementation}
\label{app:Num_Impl}

We discuss the convergence criteria and the numerical implementation of an algorithm that can reach the thermodynamic limit when the antiferromagnetic correlation length diverges. 

\subsection{Convergence Criteria}

We used the following criteria for  convergence: a small relative change of the parameter $\delta=1-U_{sp}/U_{crit}$ and a small relative change of  $U_{crit}=2/\chi^{(2)}(q_{max}, 0)$. 
Both criteria have to be satisfied but, in practice, the second criterion is always satisfied when the $\delta$-criterion is satisfied. 
The $\delta$-criterion is a strong criterion because  $\delta$ is the smallest parameter in the problem and we are looking at a small relative change of this parameter. 
It is also directly related to physical quantities: the correlation length $\xi$ and spin susceptibility $\chi_{sp}(q_{max}, 0)$. 
In addition, we have calculated the relative change of the Frobenius norm $ ||\mathcal{G}_{fb}^{(m+1)} - \mathcal{G}_{fb}^{(m)}||/||\mathcal{G}_{fb}^{(m)}||$ where $\mathcal{G}_{fb} $ is the feedback Green function used in the calculation of $\chi^{(2)}$. 
We checked that it is always small at the convergence point. 
Note that the final $\Sigma$ and $\mathcal{G}$ are calculated only once after convergence is achieved.
	
For each iteration $m$ of the Green function $\mathcal{G}_{fb}^{(m)} $. the renormalized interaction $U_{sp}$ is found numerically using the self-consistency conditions Eqs.~\ref{eq:chispqiqn}, \ref{eq:sumrule_sp} and \ref{eq:ansatz_tpsc}. 
At low temperatures, this step  requires extra care because the denominator of the spin susceptibility $\delta=1-U_{sp}\chi^{(2)}(q_{max}, 0)/2$ becomes exponentially small. 
The value of $\delta$ must be real because $\chi^{(2)}$ in Matsubara frequencies is real by construction, as can be checked from \eref{eq:chi2}. 
However, due to the finite precision, there is very small imaginary part of  $\chi^{(2)}(q_{max}, 0)$ and it can interfere with the self-consistent solution for  $U_{sp}$. 
It thus important to make sure that only the real part of $\chi^{(2)}(q, iq_n)$ is used in the calculation.  

\subsection{Thermodynamic Limit}
 When the correlation length grows exponentially in the renormalized-classical regime, the calculation of the sum over $\mathbf{q}$ in the self-consistency condition \eref{eq:sumrule_sp} is challenging. 
 The required lattice size grows with the correlation length. That quickly becomes unmanageable. 
 To deal with this problem in the thermodynamic limit, we converted the sum to the integral and made changes of variables. 
 Below we describe this approach for the commensurate case $\mathbf{q}_{max}=\mathbf{Q}=(\pi, \pi)$ in some detail.  

 First, we note that in renormalized-classical regime, $\chi_{sp}(\mathbf{q},0)$ is strongly peaked at the vector $\mathbf{Q}$ so that the best change of variables to take may be infered analytically by replacing $\chi_{sp}(\mathbf{q},0)$ by its asymptotic expression (Ornstein-Zernike susceptibility):   
%%%%%%%%%%%%%%%%%  
\begin{equation} \label{eq:OZ_s}
\chi_{sp}(\mathbf{q},0) \propto \frac{1}{\xi^{-2} +(\mathbf{q}-\mathbf{Q})^2}.
\end{equation} 
%%%%%%%%%%%%
The 2D integral of this expression has a quasisingularity at $\mathbf{q}=\mathbf{Q}$ and thus equally-spaced integration routines do not work. 
Since the quasisingularity has circular symmetry, we convert the integral over $\mathbf{q}$ to cylindrical coordinates relative to $\mathbf{q}=\mathbf{Q}$:
%%%%%%%
\begin{align}
   I=\int d^2q \chi_{sp}(\mathbf{q},0)=4 \int_0^{\pi/4} d \phi \int_0^{q_c(\phi)} q d q \chi_{sp}(\mathbf{q},0).
  \label{eq:int_chisp_q}	
 \end{align} 
%%%%%%	
	Here $q_c(\phi)=\pi/\cos(\phi)$ takes into account the square Brillouin zone.  
 The integral over angle is unremarkable and can be done using standard integration routine. 
 For the integral over $q$ we make the change of variables $\ln (q^2 +\xi^{-2})=x$. The integral then becomes:
\begin{align}
   I=2 \int_0^{\pi/4} d \phi \int_{-2\ln \xi}^{\ln(q_c^{2}(\phi) + \xi^{-2})} d x e^{x}   \chi_{sp}(\mathbf{q}(x,\phi),0)
  \label{eq:int_chisp_x}	
\end{align}
The integrand in this integral does not have a quasisingularity at the lower limit and it can be evaluated using standard integration routines. 
Note, that the number of function evaluations increases with $\xi$ but only logarithmically, which effectively solves the problem. 

To calculate the integral $I$ numerically, we need to evaluate the function  $ \chi_{sp}(\mathbf{q}(x,\phi),0) $ between points on a finite mesh in $q$-space. 
We did this by first interpolating a smooth function $ \chi^{(2)}(\mathbf{q}(x,\phi),0)$ and then using the RPA-like \eref{eq:chispqiqn} for $ \chi_{sp}$ with $\chi^{(1)}$ replaced by $\chi^{(2)}$. 
When interpolating $ \chi^{(2)}(\mathbf{q}(x,\phi),0)$, it is important to use spline interpolation rather than linear interpolation because the latter leads to a nonphysical finite-temperature phase transition. 
This is because $\int_0^{q_c(\phi)} q d q \frac{1}{\xi^{-2}+q} $ then has a finite limit as $\xi \rightarrow \infty$. 

We used the Python interpolation object RegularGridInterpolator with the spline option and the integration routine dblquad. 
This works well and produces the expected result: the integral $I$ increases logarithmically as $\xi \rightarrow \infty$.
Unfortunately, the spline interpolation is relatively slow. 
For this reason and another one described latter, we used the following strategy. 
 
The calculation starts and proceeds until convergence on a finite lattice (256*256). After that, the step with the infinite lattice correction is used to calculate $U_{sp}$, $\chi_{sp}$ and $\xi$. Note that the self-energy is calculated only on the finite mesh. This is because in the Matsubara frequency representation the self-energy has a finite limit when $\xi >> \xi_{th}$ (see Eqs.~\ref{eq:sEnergy_k_1}, and \ref{eq:sEnergy_st_R_M_Delta}). We also checked, explicitly, that the result for $\Sigma(k,ik_n)$ practically does not change when the lattice size doubles and quadruples and the correlation length is large. 

 There is another reason to start with a finite lattice: convergence. To find $U_{sp} $ with a root finder, we need to bracket it between  lower and upper limits. 
 Taking $U_{sp\_upper}$ slightly below $U_{crit}$ always works on a finite lattice. 
 In theory, this should also work on the infinite lattice because the integral $I$ diverges when $\xi \rightarrow \infty $. 
 However, the divergence is only logarithmic. 
 At low temperatures and when convergence starts from a TPSC calculation, the system is so far from convergence that $U_{sp\_upper}$ needs to get within $10^{-12}$ of  $U_{crit}$. 
 With such small numbers, one runs into  numerical stability issues (rounding errors when multiplying very large numbers by small ones in \eref{eq:int_chisp_x} ). 
 This is also a limiting factor to reach $\xi > 10^6 $. 
 Of course, such a large $\xi$ is mostly of academic interest because real quasi-two dimensional system will undergo a three-dimensional phase transition because of small couplings in the perpendicular direction. 
 
 However, if one is interested in the Matsubara self-energy at very low temperature to find the zero-temperature gap, one can suppress the infinite-lattice step, but allow accurate finite mesh calculations of the Matsubara self-energy, of double-occupancy and of other physical properties whose dependence on finite system size is relatively weak. 

\bibliography{Bibliography.bib}

\end{document}